# Probelab ReImager: An Open-Source Software for Streamlining Image Processing in an Electron Microscopy Laboratory

Noah Kraft,[1,2]* and Anette von der Handt[2]

[1]Department of Computer Science, University of Minnesota, Minneapolis, MN
[2]Department of Earth and Environmental Sciences, University of Minnesota, Minneapolis, MN

*kraft270@umn.edu

**Abstract:** Providing high-quality electron images and hyperspectral X-ray maps is a focus of many modern electron microscopy laboratories. Nevertheless, further image processing and annotations are often needed to prepare them for publications and reports. For multi-user facilities, accessibility to processing software can be a limitation either through license costs or availability of processing stations. Open-source software running on multiple platforms allows for post-acquisition data processing in-lab or on user-owned devices. We developed Probelab ReImager to supersede our vendor-supplied acquisition software's exportation by being efficient and highly customizable. This article describes its main features and capabilities.

## Introduction

Most commercial vendor-supplied software excels in the acquisition and processing of data but falls short on customization and batch processing options and/or support for modern imaging formats. Export and customization of these images in third-party software is often costly, time-consuming, and adds to the workload to achieve publishable products for scientific journals. In our case, we use Thermo Scientific's Pathfinder and its predecessor NSS software to capture electron images and X-ray intensity maps on our JEOL JXA-8530FPlus electron microprobe. While this software offers data export to various report templates, it lacks a batch export option for images in their native resolution. Furthermore, it does not allow for visual customization, which is often necessary for publication or documentation.

To fill this gap, we created the program Probelab ReImager, which facilitates easy batch exportation and customization—including annotations—of electron images and X-ray intensity maps. The current software version also supports images acquired through the JEOL PC-SEM software and integration with other image-acquisition software commonly found in electron microscopy labs.

## About Probelab ReImager

Probelab ReImager was written in NodeJS, using Electron to create an easy-to-use and well-supported program, resulting in a mobile and flexible web browser-based desktop application. Probelab ReImager is currently supported and released on Windows 7, 8, and 10, and on MacOS. Distribution is through a central web domain with an application programming interface (API) that identifies new updates for static, major, minor, and bugfix releases. While NodeJS is not extremely efficient with handling large datasets such as images, it offers a large community-supported set of open-source repositories and is easier to maintain and expand than similar languages like Python. Threading of image processing allows for faster processing of images without lag or "freezing" during use. Image processing can be intense, especially for 4096-resolution images, thus an average of 300 MB of memory is required while running. On an average Intel i5 5th generation CPU, it fully processes, customizes, and exports a 4096-resolution image in less than 20 seconds, with mobile processors taking up to 30 seconds.

Multiple layers of cache are used to increase image processing and rendering. The image displayed to the user is a PNG generated by the backend process. This image is cached as layers, which allow for quick change of X-ray map layers, individual points, and scale bar position without re-rendering large portions of the image. Further development is expected to reduce the time to process images at the expense of more memory usage.

## Download, Install, and Execute

Probelab ReImager is available for Windows and Mac platforms. The installer can be downloaded from the website https://reimager.probelab.net, and instructions on how to install on various devices are provided. Some browsers may pop up additional dialogs when downloading, such as Microsoft Edge, Chrome, or Firefox. By following these dialogs and running the installer, Probelab ReImager will be available at any time. Probelab ReImager checks online when the internet is connected and gives a visual notification if updates are available. Users can elect to update if necessary.

## File Handling and Exporting

Probelab ReImager allows loading of any project directory created through NSS or Pathfinder in order to view the images. It supports Samba and other network storage solutions with the native user interface. This makes it highly flexible and portable, as project data can be moved or copied from the original acquisition location and still accessed. Equally, any directory and file name can be chosen when saving the images individually or by batch exporting them all to the same directory automatically using a custom naming scheme.

Data types from Pathfinder/NSS that are currently supported are Point ID/Point & Shoot, Spectral Imaging, and Linescans. TIFF images acquired through the Pathfinder Electron Imaging Mode are incompatible due to the current data storage method used by Pathfinder but may change in future versions. Our current workflow addresses this shortcoming by acquiring all electron images in the Point ID mode and adding a single spectrum acquisition using the shortest live time setting available.

All image formats (TIFF, JPEG, BMP) acquired through the JEOL PC-SEM interface are fully supported. In the current program version, the JEOL image footer is cropped out and scale bars can be added. This allows streamlining of image representation between all image-acquisition software and provides processing through an open-source interface running on multiple operating systems.

## Graphic User Interface and Workflows

Figure 1 shows the graphic user interface (GUI) of Probelab ReImager. The interface has three main sections with data file handling and exportation options on the left (Panels [1] and [2]), image display in the center (Panel [3]), and image annotation settings on the right (Panels [4]–[6]).

When customizing and annotating the images or X-ray maps, the user has three workflows available. One can either (1) customize the image settings and export them on an image-by-image basis, (2) apply the same settings to selected images regardless of the mix of image resolutions and batch export them, or (3) save them to a processing queue to export later (Figure 1, Panel [2]). The same image can be added multiple times with different annotations or scale bar settings in the last option. Selected changes will be visualized after a manual refresh through the Refresh Image button (Figure 1, Panel [3]).

Usually, single images can be stored or exported; this acts the same as when exporting saved images and will bring up a dialog for each image exported. To facilitate batch export, images can be mass-selected by shift-clicking or ctrl-clicking multiple single images like a file browser. By doing this, all images selected can be stored in the processing queue or exported immediately. There are two modes to export in this method, changed by toggling the Easy button that appears when selecting multiple images. Alternatively, by clicking Export All, Easy mode will be activated, and all images in the project will be exported.

Easy mode brings up the save dialog once, displaying "{name}.png" as the default file name. This "{name}" will automatically be replaced with the corresponding image name from the project. For example, "{name}-1-9-2020.png" will save each image with its own name and the date behind each image. This is commonly known as variable substitution, and new variables will be introduced in the future.

When naming files, some programs limit the number or types of characters that can be used. Probelab ReImager allows for the use of any character the operating system will allow, with the maximum name length allowed. In Windows, this means that \, /, |, :, and several other special characters cannot be used.

While many programs still only export images as TIFFs, this slightly archaic image format has been superseded by formats such as PNG for lossless and JPG for lossy compression. Both of these formats offer more advanced metadata and compression and are supported on every operating system. To facilitate the change of image formats, Probelab ReImager by

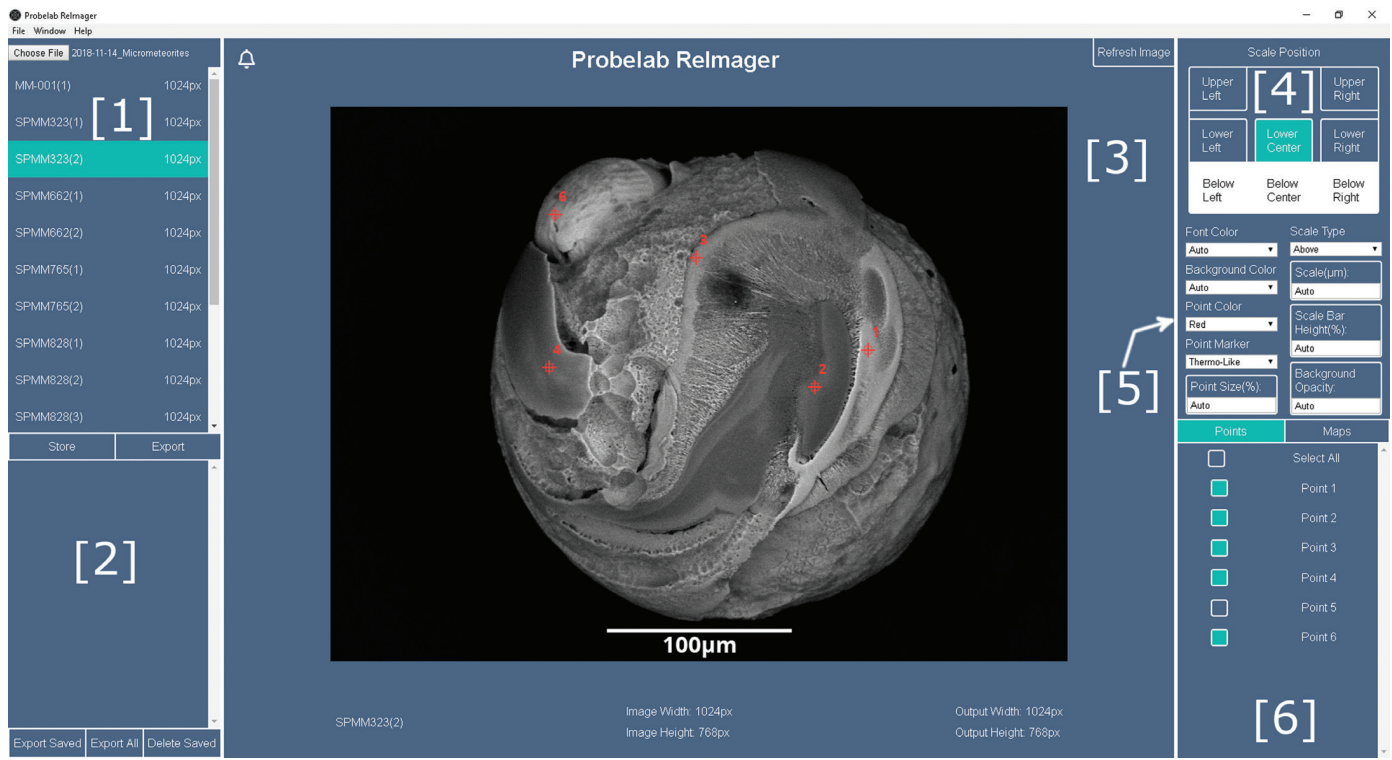

**Figure 1:** Screenshot of the GUI of Probelab ReImager. The interface has six main areas with (1) list of data files; (2) batch processing queue; (3) image preview, notification bell, and image refresh; (4) scale bar positioning; (5) annotation and scale bar customization settings; and (6) point and map selection.

Microscopy TODAY **39**

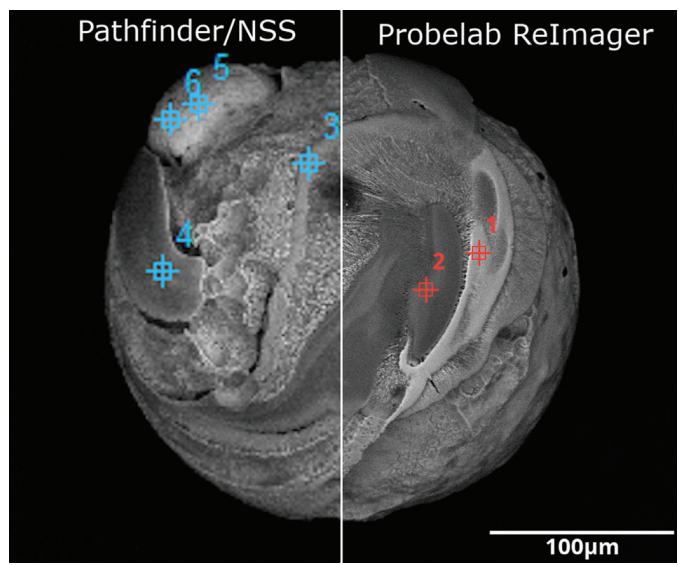

**Figure 2:** Comparison of the same image exported through the Pathfinder/NSS software (left) and through Probelab ReImager (right). The left image export is more pixelated because the software down-samples exported images relative to their native resolution. Also, the interface does not provide any customization of location markers as shown for the spectra acquisitions. Probelab ReImager exports in native resolution without down-sampling.

default exports images to PNG but also supports JPG, WebP, and TIFF.

Most importantly, Probelab ReImager always exports the customized image in the acquired resolution. Pathfinder/NSS either exports the raw image or the annotated but down-sampled image as shown in a direct comparison between both outputs in Figure 2.

## Scale Bar Calibration

One of the most essential applications in electron microscopy is the sizing and imaging of small features. Instrument scale bars have to be calibrated initially and checked periodically. Probelab ReImager comes with a default calibration acquired on the University of Minnesota's JEOL JXA-8530FPlus electron microprobe.

The first step for a user or a lab is to calibrate Probelab ReImager's scalebar. While the calculated constant used for the Probelab's Microprobe should be close for many electron microprobes, it may differ by manufacturer, device calibration, and the individual device used to take images and data.

The process to calibrate the scale bar is a straightforward trend estimation of the pixel size at various magnifications (Table 1 and Figure 3). By calibrating the scale bar, Probelab ReImager can recreate the scale bar more accurately than some vendor software and allow dynamic changes of scale bar length while maintaining accuracy. It is recommended that this be updated when the device is recalibrated during maintenance, though a well-maintained instrument may not see much, if any, drift.

## Image Annotation: Scale Bar Locations and Appearance

Both electron and X-ray map images provide full customization of the relative position of scale bars and its appearance, as this is often desired for publication-ready documentation. Probelab ReImager offers eight different scale bar locations, five on the image itself and three on a data bar below the image (Figure 1, Panel [4]). For locations on the image itself, the user can add a black or white background with custom opacity (Figure 1, Panel [5]) to improve scale bar visibility without obscuring potentially essential image features.

Scale bar size can either be automated and then calculated to fall between a third and a quarter length of the image or be user-defined, for example, if a consistent scale bar length is wanted for comparison across image magnifications (Figure 4).

Font color can be either black or white, and font size is defined as a percentage of the image resolution. Scale bar height can also be customized and defined as a percentage of the text height. The bar and text position can be swapped to place the text above or below the bar.

## Image Annotation: Spectra Location Visualization

Pathfinder/NSS allows acquisition of spectra as points, rectangles, circles, or polygons with documentation of their relative location. Probelab ReImager supports all of these acquisition types and expands on their customization options (Figure 1, Panel [5]). For point data, the user can choose between four different colors and four different marker types (Figure 4). Their relative size can be user-defined as a percentage of the image size. Spectra location data can be either selected individually to be displayed on an image or enabled/disabled

**Table 1:** A trend estimation of the pixel size at various magnifications to calibrate the scale bar.

| Magnification | X (µm) | Y (µm) | Diagonal | Image Resolution | Pixel Size (µm) |
|---|---|---|---|---|---|
| 40 | 2988.281 | 2146.875 | 3679.524 | 1024 | 2.918243164 |
| 100 | 1195.313 | 858.75 | 1471.81 | 1024 | 1.167297852 |
| 250 | 478.125 | 343.5 | 588.724 | 1024 | 0.466918945 |
| 500 | 239.063 | 171.75 | 294.362 | 1024 | 0.233459961 |
| 1000 | 119.531 | 85.875 | 147.181 | 1024 | 0.116729492 |
| 2000 | 59.766 | 42.937 | 73.59 | 1024 | 0.058365234 |
| 4000 | 29.883 | 21.469 | 36.795 | 1024 | 0.029182617 |

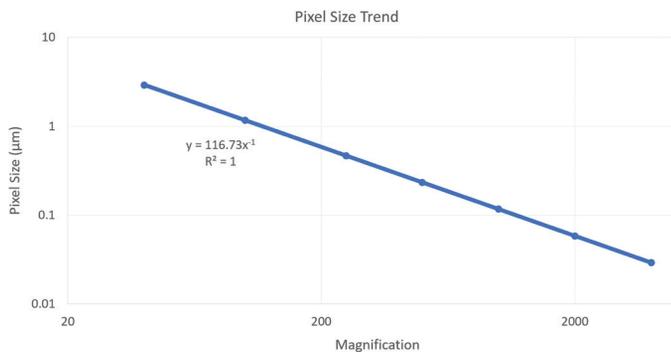

**Figure 3:** Scale bars are calibrated by calculating the pixel size for a standard image resolution of 1024 pixels over the range of magnification available. The resulting power-law trend fit is stored as the pixel size constant (here: 116.73) in the Probelab ReImager settings.

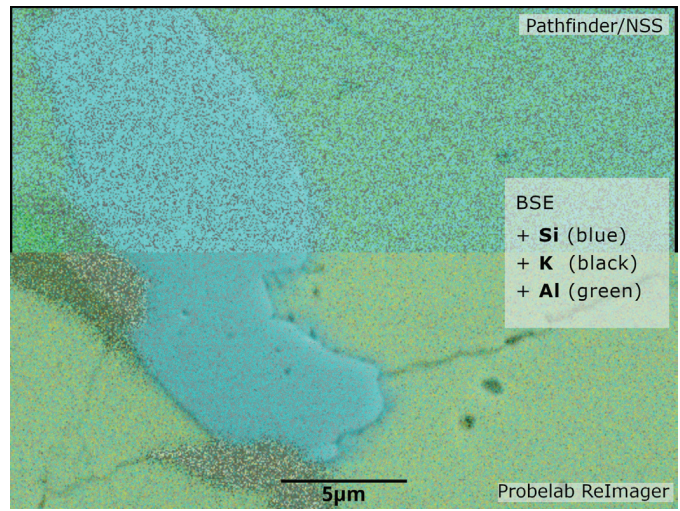

**Figure 5:** Comparison between outputs from Pathfinder/NSS (top) and Probelab ReImager (bottom). In both cases, X-ray maps were processed previously in Pathfinder with the Hot Pixel Suppression filter and manual contrast of K. Probelab ReImager's opacity options improve the noise of the raw map output and image clarity. In addition, output from Pathfinder adds a thin black edge around the maps.

globally for all available positions across images for batch export purposes (Figure 1, Panel [6]).

## Image Annotation: X-ray Intensity Maps Visualization

Pathfinder/NSS offers hyperspectral imaging, and an impressive library of X-ray map processing and filtering features is available. Raw X-ray intensity maps can be processed using Pathfinder or NSS's Counts, Quant, COMPASS (Principal Component Analysis), and Phases routines and exported as individual TIFF files. These can then be layered and further annotated using Probelab ReImager. This allows for the traditional and well-tested map processing in Pathfinder and NSS to integrate seamlessly with Probelab ReImager.

Notably, Probelab Reimager adds the option of defining layer order and opacity. Figure 5 shows a comparison of the raw, unquantified Pathfinder/NSS X-ray maps on top using only their custom contrast option; the bottom half of Figure 5 shows the same three maps but using Probelab ReImager's opacity system.

Pathfinder/NSS allows for custom contrasting of element X-ray maps. These maps are then layered on top of one another with no opacity. Probelab ReImager takes an alternative approach where each map is given the same custom opacity and then layered, minimizing the noise of raw map output and providing a more useful overview image. The layer order can be changed by drag-and-drop. Opacity mixes multiple colors on the same pixel if each element is present, allowing for easier visual identification of changes. In addition, Probelab ReImager allows use of a solid color background instead of the electron image (Figure 1, Panel [6]).

## Access to Probelab ReImager

Probelab ReImager has been released to the public and is available at https://reimager.probelab.net. While currently limited to data acquired by Thermo Scientific's Pathfinder/NSS software and JEOL EPMA and SEM-platform software PC-SEM, we are presently exploring the support of images acquired through Cameca's PeakSight and Probe Software Inc.'s Probe for EPMA software. Support for Bruker, Oxford, and other software vendors is anticipated. Code is available on Github through the webpage and licensed under MIT copyright license.

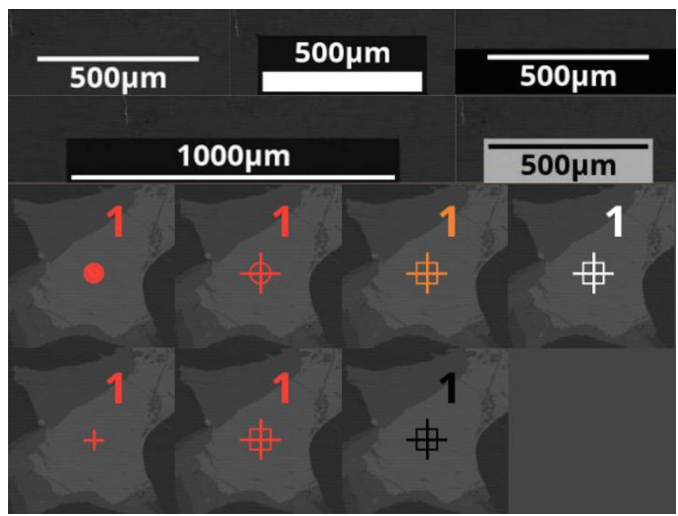

**Figure 4:** Examples of scale bar (top) and spectra location marker (bottom) appearances in Probelab ReImager. Scale bar width, length, and annotation can be modified, and the user can choose between four different colors and four different markers for point spectra locations.

## Acknowledgements

We would like to acknowledge support through NSF grants #1849465, #1625422 as well as Prof. Marc Hirschmann (UMN). We also thank Dr. Steve Seddio (ThermoFisher Scientific) for helpful discussions and guidance.